# Stability of small amplitude normal modes of a Bose-Einstein condensate with a singly quantized vortex confined in an optical lattice

Aranya B Bhattacherjee*, O. Morsch and E. Arimondo

INFM, Dipartimento di Fisica E.Fermi, Universita di Pisa, Via Buonarroti 2, I-56127, Pisa, Italy

**Abstract:** We study the dynamics of a BEC with a singly quantized vortex, placed in the combined potential of a 1-D (2-D) optical lattice and an axi-symmetric harmonic trap. A time-dependent variational Lagrangian analysis shows that an optical lattice helps to stabilize the vortex which in absence of the optical lattice is unstable. We find that the normal modes are stable only if the depth of the optical potential is more than a certain critical value. This critical value of the optical potential depends on the 2-D interaction parameter. In general higher the interaction parameter, lower the value of the optical potential required to stabilize the vortex. The BEC with the singly quantized vortex is found to be relatively more unstable in a 2-D optical lattice compared to a 1-D optical lattice.

*Permanent Institute: Department of Physics, Atma Ram Sanatan Dharma College, University of Delhi (South campus), Dhaula Kuan, New Delhi-110021, India.

The optical manipulation of a Bose-Einstein condensate provides a great potential for the investigation of its properties. The theoretical research of the interaction of a BEC with light waves and related topics in the quantum statistics of ultra cold atomic ensembles has made a lot of progress [1] and has opened new opportunities to study atomic BEC in optical lattices. Some very interesting phenomena have been observed for a BEC confined in an optical lattice such as Bloch oscillations [2] and superfluid-Mott insulator transition [3]. Recently Diakonov et. al. [4] investigated theoretically the dynamics of a BEC near the boundary of the first Brillouin zone and found the remarkable fact that the current of the condensate for quasi-momenta in the vicinity of the zone boundary has a singular behavior and it was shown that the chemical potential of the condensate becomes multi-valued near the zone boundary. Recently it was proposed that Bragg reflection could provide a new mechanism for generating vortices in BECs trapped in an optical lattice [6]. The physical picture of the vortex dynamics was unclear and hence the question of vortex stability in optical lattices still remains unanswered. Therefore an especially important and intriguing question concerns the stability of a condensate with a vortex confined in an optical lattice. Martikainen et al. [7] have investigated the transverse and longitudinal excitations of a stack of weakly-coupled two-dimensional BEC's that is formed by a one-dimensional optical lattice. They considered a system which contained a vortex along the direction of the lasers creating the optical lattice. Using a variational approach, they have determined dispersion relations of the monopole and quadrapole modes. Here, we use a variational Lagrangian procedure (similar to the one used in Ref.[8] to study the dynamical normal modes of a vortex and a condensate in a rotating axi-symmetric trap) to provide a simple, direct and intuitive treatment of the small amplitude normal modes of a small BEC containing a singly quantized vortex in a direction *perpendicular* to the optical lattice. In particular, we assume that the vortex is created in the BEC by known techniques [9-13] and then the optical lattice is slowly ramped up, placing the BEC in the optical lattice without any induced linear velocity.

We consider a N-particle BEC containing a singly quantized vortex (in the $z$ direction and slightly off-center) in a 1-D optical lattice and a 3-D harmonic trap. The 1-D optical lattice is of the form $V_{OL}^{1D} = V_O \cos^2(\frac{\pi x}{d})$, where $d$ is the period of the lattice. The trap frequency along the $z$-direction is assumed to be high enough for the BEC dynamics to be reduced to 2-D. The potential is then given by $V(x,y) = V_{OL}^{1D}(x) + \frac{1}{2}m\omega_\perp^2(x^2 + y^2)$. Where $m$ is the mass of a single atom, $\omega_\perp$ is the frequencies of the harmonic trap along the radial direction and characterizes the symmetry about the $z$-axis. For such a general system the dimensionless Lagrangian is given by:

$$L(\psi) = \int_{-\infty}^{\infty} dy \int_{-\infty}^{\infty} dx \left[ \frac{i}{2}\left(\psi^* \frac{\partial \psi}{\partial t} - \psi \frac{\partial \psi^*}{\partial t}\right) - \psi^* H_O \psi - \psi^* V_{OL} \psi - \gamma_{2D}|\psi|^4 \right] \quad (1)$$

Where $H_O = \frac{k}{2}[-\frac{\partial^2}{\partial x^2} - \frac{\partial^2}{\partial y^2} + (x^2 + y^2)]$ is the Hamiltonian of the non-interacting condensate. All quantities are expressed in dimensionless units. The wave function is normalized to the particle number $N$. All energies are scaled with respect to the recoil energy $(E_{rec}) = \hbar^2 k_L^2 / 2m$, $k_L$ is the wave number of the two counter-propagating lasers required to create the optical lattice. The coordinates are scaled with the radial oscillator length $a_\perp = \sqrt{\hbar / m\omega_\perp}$. $\gamma_{2D} = \sqrt{8\pi}\hbar^2 a / a_z a_\perp^2 E_{rec} m$ is the 2D dimensionless interaction parameter related to the s-wave scattering length $a$ and $a_z = \sqrt{\hbar / m\omega_z}$ is the axial oscillator length. $k$ is the dimensionless parameter related to the frequency of the harmonic trap along the radial direction as $k = \omega_\perp / \omega_{rec}$. In the present application, we use a variational trial wave function with time dependent parameters. The resulting Lagrangian will contain terms that involve the first time derivatives of these parameters along with the negative of the variational Gross-Pitaevskii Hamiltonian.

We are interested in the motion of a vortex relative to the centre of mass of the condensate, so it is natural to take a time-dependent trial function in the form

$$\psi(x,y,t) = C[\vec{r}_o(t)\vec{r}_1(t)] f[\vec{r}-\vec{r}_0(t)] F[\vec{r}-\vec{r}_1(t)] \qquad (2)$$

Where $C[\vec{r}_o(t)\vec{r}_1(t)]$ is the normalization constant. The function $f[\vec{r}-\vec{r}_o(t)]$ characterizes the vortex inside the trap and has the form $[(x-x_0(t))+i(y-y_0(t))]$. The function $F[\vec{r}-\vec{r}_1(t)]$ describes the condensate density distribution and has the form $F_x(x-x_1)F_y(y-y_1)$. The condensate density is Gaussian along the $y$-direction, hence $F_y(y-y_1) = \exp\{-\frac{1}{2}W_\perp^2[y-y_1(t)]^2\}$. The optical potential is assumed to be weak so that there is no fragmentation of the condensate. Consequently we assume that the optical potential weakly perturbs the Gaussian wave function along the x-direction, hence $F_x(x-x_1) = \exp\{-\frac{1}{2}W_\perp^2[x-x_1(t)]^2\}\{1-\alpha\cos\pi\beta[x-x_1(t)]\}$. Where $\alpha = V_o/\mu_{2D}$, $\alpha < 1$, The chemical potential $\mu_{2D}/E_{rec} = \frac{\hbar\omega_z}{2\omega_{rec}} + \gamma_{2D} n_{2D}$ for $\omega_z \gg \omega_\perp$ [14], $n_{2D}$ is the dimensionless 2D density of the cloud and $\beta = a_\perp/d$. $W_\perp = a_\perp/b_\perp$, $b_\perp$ is the 2D width of the BEC along the radial direction taking into account interactions. $b_\perp = \left\{\left(\frac{2}{\pi}\right)^{1/2}\left(\frac{Na}{a_z}\right)+1\right\}^{1/4} a_\perp$, found by minimizing the unperturbed Lagrangian. The choice of the particular part of the trial wavefunction modulated by the weak optical potential is motivated by the fact that, $|\psi_{OL}^{1D}(x)| \propto \left(1-\frac{V_{OL}^{1D}}{\mu_{2D}}\right)^{1/2}$. Finally $[x_0(t), y_0(t)]$ are the displacements of the vortex core position while $[x_1(t), y_1(t)]$ are the displacements of the BEC center of mass. The trial wave function for the 1-D optical lattice has the form

$$\psi_{1D}(x,y,t) = C[\vec{r}_0(t),\vec{r}_1(t)]\{[x-x_0(t)]+i[y-y_0(t)]\}\exp\{-\tfrac{1}{2}W_\perp^2[x-x_1(t)]^2\}\{1-\alpha\cos\pi\beta[x-x_1(t)]\}$$
$$\times \exp\{-\tfrac{1}{2}W_\perp^2[y-y_1(t)]^2\}$$

(4)

In evaluating the integration in $L(\psi)$, we retain all terms up to second order in the small time-dependent parameters. The resulting effective Lagrangian takes a simple form in terms of two new parameters $\vec{\delta} = (\vec{r}_1 - \vec{r}_0)$ and $\vec{\varepsilon} = (2\vec{r}_1 - \vec{r}_0)$

$$L_{1D} = \frac{N}{1+2\delta^2}\left\{2(\delta_x\dot{\varepsilon}_y - \delta_y\dot{\varepsilon}_x + 2\delta_y\dot{\delta}_x - 2\delta_x\dot{\delta}_y) - 2W_\perp^2 k(1+\delta^2) - \frac{k}{2W_\perp^2}(1+\varepsilon^2)\right\} - \Gamma_{1D}\frac{(1+8\delta^2)}{(1+2\delta^2)^2}$$
$$- \frac{4N\pi^2\alpha^2\beta^2 k}{3(1+\alpha^2/2)W_\perp^2} - \frac{V_o N(4+3\alpha^2)}{8(1+\alpha^2/2)}$$

(5)

Where $\delta^2 = \delta_x^2 + \delta_y^2$, $\varepsilon^2 = \varepsilon_x^2 + \varepsilon_y^2$, $\Gamma_{1D} = \frac{\gamma_{2D} N^2(1+3\alpha^2)}{2\pi(2+\alpha^2)^2 W_\perp^2}$. In deriving the above Lagrangian we have retained $\alpha$ up to second order and assumed $\beta/W_\perp \gg 1$ for typical experimental situations. The equations of motion for $\vec{\delta}$ and $\vec{\varepsilon}$ are found from the Lagrange equations:

$$\frac{d}{dt}\left(\frac{\partial L}{\partial \dot{X}_i}\right) - \frac{\partial L}{\partial X_i} = 0, X = \delta,\varepsilon, i = x,y \tag{6}$$

The effective Lagrangian eqn. (5) yields four coupled first-order differential equations for the displacements $\vec{\delta}$ and $\vec{\varepsilon}$. Eliminating the latter through the Lagrange equations $\varepsilon_x = \frac{2W_\perp^2\dot{\delta}_y}{k}$ and $\varepsilon_y = \frac{-2W_\perp^2\dot{\delta}_x}{k}$ leads to two coupled equations for the displacements $\delta_x$ and $\delta_y$.

$$\ddot{\delta}_x + \lambda_{1D}\delta_x + \frac{3}{2}\frac{k}{W_\perp^2}\dot{\delta}_y = 0 \quad \text{and} \quad \ddot{\delta}_y + \lambda_{1D}\delta_y - \frac{3}{2}\frac{k}{W_\perp^2}\dot{\delta}_x = 0 \tag{7}$$

Where $\lambda_{1D} = \frac{k}{W_\perp^2}\left(\frac{2\Gamma_{1D}}{N} - \frac{k}{2W_\perp^2} - kW_\perp^2\right)$. This involves uniform oscillatory motion for certain values of $\lambda_{1D}$ and $k$, as can be seen by considering the complex variable $\delta_x(t) + i\delta_y(t) = \delta(t)$ that obeys the second order differential equation

$$\ddot{\delta}(t) - \frac{3}{2}i\frac{k}{W_\perp^2}\dot{\delta}(t) + \lambda_{1D}\delta(t) = 0. \tag{8}$$

The general solutions of equation (8) are well known and represents a sinusoidal motion for a characteristic positive frequency

$$\omega_{1D} = \sqrt{k^2\left\{8\sqrt{\frac{2}{\pi}}\left(\frac{Na}{a_z}\right)f(\alpha)\left(\sqrt{\frac{2}{\pi}}\left(\frac{Na}{a_z}\right)+1\right) + \frac{1}{4}\left(\sqrt{\frac{2}{\pi}}\left(\frac{Na}{a_z}\right)+1\right) - 4\right\}} \tag{9}$$

Where, $f(\alpha) = \frac{(1+3\alpha^2)}{(2+\alpha^2)^2}$. The existence of imaginary $\omega_{1D}$ implies instability of the $\delta$ modes. For $\omega_{1D}^2 < 0$, the modes have exponentially decaying solutions as well as exponentially growing solutions. Eventually, the exponentially growing solution will dominate. When this occurs the vortex moves out of the condensate.

We also performed the variational Lagrangian procedure on a BEC with a singly quantized vortex in a symmetric 2-D optical lattice of the form $V_{OL}^{2D} = V_O\{\cos^2(\frac{\pi x}{d}) + \cos^2(\frac{\pi y}{d})\}$. We define a symmetric 2-D optical lattice as one in which the 1-D optical lattices along $x$ and the $y$ direction are same with respect to lattice depth and periodicity. The trial wave function was chosen as

$$\psi_{2D}(x,y,t) = C[\vec{r}_0(t),\vec{r}_1(t)]\{[x-x_0(t)]+i[y-y_0(t)]\}\exp\{-\tfrac{1}{2}W_\perp^2[x-x_1(t)]^2\}\{1-\alpha\cos\pi\beta[x-x_1(t)]\}$$
$$\times \exp\{-\tfrac{1}{2}W_\perp^2[y-y_1(t)]^2\}\{1-\alpha\cos\pi\beta[y-y_1(t)]\}$$

(10)

In the usual manner we find the frequency for the 2-D optical lattice configuration as

$$\omega_{2D} = \sqrt{k^2\left\{8\sqrt{\frac{2}{\pi}}\left(\frac{Na}{a_z}\right)f^2(\alpha)\left(\sqrt{\frac{2}{\pi}}\left(\frac{Na}{a_z}\right)+1\right)+\frac{1}{4}\left(\sqrt{\frac{2}{\pi}}\left(\frac{Na}{a_z}\right)+1\right)-4\right\}}$$

(11)

Figure 1 shows a comparative plot of $\omega_{1D}$ and $\omega_{2D}$, as a function of lattice depth. For $Na/a_z = 1.75$ reveals that $\omega_{2D}^2 < 0$ (shown as W2 (dashed line) for $V_o \leq 0.04 E_{rec}$. This indicates instability and hence the 2-D optical lattice configuration is not able to maintain a stable vortex. On the other hand, for the given value of $Na/a_z = 1.75$, $\omega_{1D}^2$ is positive for the entire range of lattice depth shown by W1(solid line). The normal modes for the 2-D optical lattice could be stabilized by increasing the interaction parameter (by increasing the scattering length using Feshbach resonance or by increasing the axial frequency $\omega_z$). Increasing the parameter $Na/a_z = 2.75$, makes $\omega_{2D}^2$ positive for all values of the optical depth (shown as a dotted line (W3). A numerical study on a configuration similar to this work was done recently by Kevrekidis et. al. [15]. They considered much higher lattice depths ($0.5 E_{rec}$) and found that the lattice depth had little effect on the vortex dynamics. In summary we have studied, using the time-dependent Lagrangian approach, the dynamics of a BEC in a 1-D and 2-D optical lattice with a singly quantized vortex line perpendicular to the optical lattice. We find that the optical lattice is a new experimental tool to stabilize a vortex. The normal modes are stable only if the depth of the optical potential is more than a certain critical value. This critical value of the optical potential depends on the 2-D interaction parameter. Higher the interaction parameter, lower the value of the optical depth required to stabilize the vortex. The BEC with the singly quantized vortex is found to be relatively more unstable in a 2-D optical lattice compared to a 1-D optical lattice. We find that for a 2-D optical potential the minimum lattice depth above which the vortex is stable is higher than that for a 1-D optical potential configuration. These results are important from an experimental point of view as they provide an estimate of the experimental parameters that are needed to have a stable vortex in an optical lattice.

**Acknowledgements:** A.Bhattacherjee acknowledges support by the Abdus Salam International Centre for Theoretical Physics, Trieste, Italy under the ICTP-TRIL fellowship scheme.

**FIGURE CAPTIONS**

Figure 1. A plot of $\omega_{1D}$ (W1(x)) and $\omega_{2D}$ (W2(x)), as a function of lattice depth $V_o$ in units of $E_{rec}$ for $Na/a_z = 1.75$, $\omega_z = 2\pi \times 150\,\text{Hz}$, $\omega_\perp = 2\pi \times 20\,\text{Hz}$, $\omega_{rec} = 20\,\text{kHz}$. The dotted line (W3(x)) is a plot of $\omega_{2D}$ for $Na/a_z = 2.75$. The expression for the chemical potential $\mu_{2D}/E_{rec} = \omega_z/2\omega_{rec} + 8k\left(\sqrt{\frac{2}{\pi}} Na/a_z\right)^{1/2}$ has been used to plot this figure.

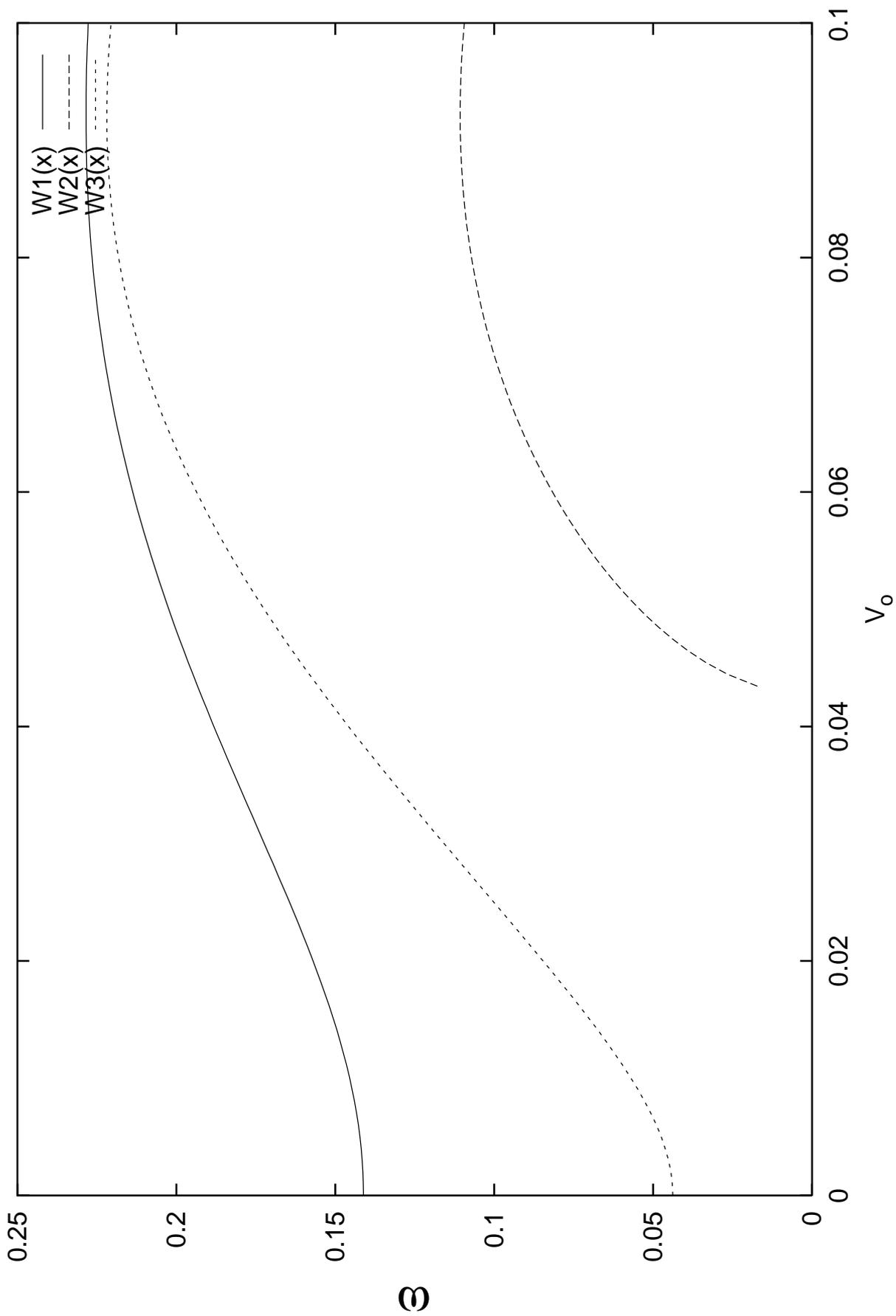

Figure 1